\documentclass[twocolumn,superscriptaddress, floatfix,amsmath,amssymb,nofootinbib]{revtex4-1}
\usepackage[english]{babel}
\usepackage{graphicx}
\usepackage{dcolumn}
\usepackage{bm}
\usepackage{verbatim}
\usepackage{mathrsfs}
\usepackage{natbib}
\usepackage{color}
\usepackage{hyperref}
\usepackage[dvipsnames]{xcolor}
\usepackage[utf8]{inputenc}
\pdfoutput=1

\newcommand{\tcr}{\textcolor{red}}
\newcommand{\tcb}{\textcolor{blue}}

\newcommand{\be}{\begin{equation}}
\newcommand{\ee}{\end{equation}}

\begin{document}
\title{Over the horizon: distinguishing the Schwarzschild spacetime and the $\mathbb{RP}^3$ spacetime using an Unruh-DeWitt detector}
\author{Keith K. Ng}
\affiliation{Department of Physics and Astronomy, University of Waterloo, Waterloo, ON, N2L 3G1, Canada}
\author{Robert B. Mann}
\affiliation{Department of Physics and Astronomy, University of Waterloo, Waterloo, ON, N2L 3G1, Canada}
\affiliation{Institute for Quantum Computing, University of Waterloo, Waterloo, Ontario, N2L 3G1, Canada}
\affiliation{Perimeter Institute for Theoretical Physics, Waterloo, ON, N2L 2Y5, Canada}
\author{Eduardo Mart\'{i}n-Mart\'{i}nez}
\affiliation{Institute for Quantum Computing, University of Waterloo, Waterloo, Ontario, N2L 3G1, Canada}
\affiliation{Perimeter Institute for Theoretical Physics, Waterloo, ON, N2L 2Y5, Canada}
\affiliation{Department of Applied Mathematics, University of Waterloo, Waterloo, ON, N2L 3G1, Canada}

\begin{abstract}
We show that a particle detector can distinguish the $\mathbb{RP}^3$ geon from the Schwarzschild black hole, even though they differ only by a topological identification beyond the event horizon. This shows that the detector can read out information about the non-local structure even when separated from the non-locality by an event horizon. Our analysis of the dependence of the transition on the detector  gap is novel, and in principle presents an  interesting observational signal.
\end{abstract}

\maketitle

\section{Introduction}

Black hole radiation represents the intersection of the two foundational theories in physics: general relativity and quantum physics. Hawking \citep{Hawking:1974rv} showed using quantum mechanical arguments that a black hole must emit particles, and eventually evaporate. However, even from the beginning, this was cause for some concern: Hawking radiation is purely thermal, so it cannot carry any information about the contents of a black hole. If the black hole eventually evaporates, then information about what fell into the black hole is lost; this represents a violation of quantum theory. Many papers have been published seeking to resolve this paradox (e.g. \citep{Almheiri:2012rt,Hawking:2016msc}), for it is believed that the solution may help uncover a quantum theory of gravity.

Recently considerable progress in research has taken place on another intersection of general relativity and quantum physics: namely, relativistic quantum information, which is concerned with how relativistic effects modify quantum information tasks.  A key tool in such investigations is the Unruh-DeWitt detector \citep{Unruh:1976db}, which allows us to investigate what an observer `sees', operationally. Originally used to study the Hawking radiation of a black hole, it has recently been fruitfully employed to explore such phenomena as Unruh radiation due to acceleration \citep{Unruh:1976db}, vacuum entanglement harvesting \citep{Pozas-Kerstjens:2015gta}, the effect of a local gravitational field on the vacuum \citep{Louko:2007mu}, the effect of black holes on detector response \cite{Hodgkinson:2014iua,Ng:2014kha}
the topology of flat spacetime \cite{Martin-Martinez:2015qwa}, and topological features hidden behind an event horizon \citep{Smith:2013zqa}. The Unruh-DeWitt detector also captures the fundamental features of the physics of atoms coupled dipolarly to the electromagnetic field when the exchange of angular momentum is not relevant \cite{Pozas-Kerstjens:2016rsh,MartinMartinez:2012th,Alhambra:2013uja}.
 
The observation of spacetime topology may seem to be at odds with the topological censorship theorem \citep{Friedman:1993ty}, which states that any worldline from the asymptotic past to the asymptotic future must be topologically trivial. In essence, any `wormhole' that might form collapses too quickly to be detected. However a loophole exists:  \textit{passive} detection is still possible. While topological features must be hidden behind an event horizon, a feature behind a \textit{past} event horizon could emit radiation, and therefore be detected.

Recent results concerning the quantum vacuum have lead us to examine this scenario more closely. We recently demonstrated that the quantum vacuum is sensitive to  distant gravitational fields, and that this difference in vacua was observable by an Unruh-DeWitt (UDW) detector: it can read out information about the non-local structure of spacetime even when switched on for scales much shorter than the characteristic scale of the non-locality \citep{Ng:2016hzn}. In conjunction with previous work done in $(2+1)$ dimensions \citep{Smith:2013zqa}, 
it is natural to investigate whether such detectors can be employed  to discern some information about the inner structure of a black hole, in particular its topology. This would represent a demonstration of passively detecting a classically inaccessible topological feature.  

In order to carry out such a study,  we seek to compare the Schwarzschild  spacetime to some other spacetime that  is identical everywhere outside the black hole but not inside.  Fortunately, such a solution exists: the $\mathbb{RP}^3$ geon, which differs from the Schwarzschild black hole only by a topological identification behind the horizon\footnote{In fact, this particular spacetime was mentioned by name in \citep{Friedman:1993ty} as a possibility.} While this identification
causes the geon to become subtly time dependent (e.g. only one of the Schwarzschild constant time surfaces is `smooth' inside), from a classical perspective  (i.e. a Boulware-type vacuum) this cannot be determined from outside by any classical means. Therefore, any difference must be due to quantum effects.

The use of this geon also has practical implications for carrying out the computation. Since it only varies from the Schwarzschild solution by an involution, we can use standard source-image arguments to quickly express the Wightman function of the geon in terms of that of the Schwarzschild solution. As we will discuss, this greatly shortens our calculations, and aids interpretation.

The case of the geon, as a topological identification of a black hole, was previously analyzed in $(2+1)$ dimensions by \citep{Smith:2013zqa}, demonstrating that a detector could distinguish a geon from the BTZ black hole. The field theory of the $(3+1)$-dimensional case has also been explored \citep{Louko:1998dj}  using a slightly different formalism than the one we employ. However, our paper is the first to investigate the $\mathbb{RP}^3$ geon \textit{numerically}, using a detector.  Our calculation of the transition rate of the detector is consistent with previous results in BTZ \citep{Smith:2013zqa}, while our calculation of the transition probability's dependence on radius and time matches the qualitative predictions of \citep{Louko:1998dj}. Moreover, our analysis of the dependence of the transition on the detector \textit{gap} is novel, and presents an interesting observational signal.

In this paper, we will demonstrate that although the $\mathbb{RP}^3$ geon and the Schwarzschild solution are classically identical everywhere outside the black hole, their vacua differ significantly, and can easily be observed from outside using a UDW detector. We will also demonstrate the time dependence of this vacuum, and discuss how time translations affect the result. This demonstrates that a local observer can determine the otherwise hidden topology of spacetime, at least in the context of black holes.

\section{The Unruh-DeWitt detector}

The UDW detector is a simple model of a 2-level detector locally interacting with a scalar quantum field. As mentioned above, despite its simplicity, it has been shown that this model captures many of the features of the light-matter interaction \cite{Pozas-Kerstjens:2016rsh,MartinMartinez:2012th,Alhambra:2013uja}. Its simplicity makes it an ideal theoretical tool, and it has been used in many recent relativistic quantum information calculations   \citep{Louko:2007mu,Smith:2013zqa,Hodgkinson:2014iua,Ng:2014kha,Pozas-Kerstjens:2015gta,Ng:2016hzn}.

Suppose we have an Unruh-DeWitt detector with energy gap $\Omega$. We would like to switch it on and off, then check whether or not it has been excited (i.e. the detector `clicks'). It interacts with the field via the interaction Hamiltonian
\begin{equation}
H_I=\lambda \chi(\tau) \mu(\tau) \phi(\mathsf{x}(\tau)),
\end{equation}
where $\lambda$ is a small coupling constant, $\chi(\tau)$ is the switching function, $\mu(\tau)$ is the monopole operator of the detector, and $\phi(\mathsf{x})$ is the field operator.

It is fairly simple to show that under these conditions, the transition probability of the detector, to leading order in $\lambda$, is
\begin{align}
P=&\lambda^2 |\langle 0 | \mu(0) | 0 \rangle|^2 F(\Omega) \\
F(\Omega)=&\lim_{\epsilon\rightarrow 0}\int_{-\infty}^{\infty} d\tau' \int_{-\infty}^{\infty} d\tau'' \chi(\tau')\chi(\tau'')e^{-i \Omega (\tau'-\tau'')}\nonumber\\
&\times W_\epsilon(\tau',\tau'')
\end{align}
where $W_\epsilon(\tau',\tau'')=\langle 0| \phi(\mathsf{x}(\tau')) \phi(\mathsf{x}(\tau''))|0\rangle $ is the pullback of the Wightman function to the worldline of the detector, up to some regularization $\epsilon$.

The regularization of this expression is less simple than it may appear; while the na\"{i}ve $i\epsilon$ approach will work in some cases, it can cause unphysical results in seemingly simple situations, such as uniform acceleration. However  this expression is regularization-independent if the switching function is sufficiently smooth \citep{Louko:2007mu}.  Furthermore, divergences caused by sudden switching are independent of the state of the field: therefore, we may analyze the difference of transition rates between the geon and the Schwarzschild black hole freely, i.e. the image contribution of the geon. We will investigate both aspects further.

\section{QFT on the $\mathbb{RP}^3$ geon}

\begin{figure}[hbtp]
\centering
\includegraphics[trim={0cm, 1.5cm, 0cm, 4.cm}, clip,width=0.5\textwidth]{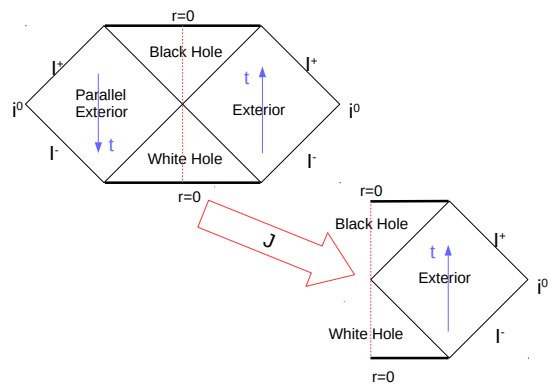}
\caption{The $\mathbb{RP}^3$ geon from the extended Schwarzschild solution.}
\label{Jfig}
\end{figure}

The $\mathbb{RP}^3$ geon differs from the Schwarzschild spacetime merely by a topological identification, as shown in figure \ref{Jfig}.
In terms of Kruskal coordinates, this map may be expressed as \citep{Louko:1998dj}
\begin{equation}
J(T,X,\theta,\phi)=(T,-X,\pi-\theta,\phi+\pi).
\end{equation}
Put in simple terms, the involution $J$ maps a point inside the horizon of a (maximally-extended) Schwarzschild black hole to an antipodal point outside the black hole. If we allow ourselves to abuse notation and use Schwarzschild coordinates on both exteriors of the black hole, this may also be written
\begin{equation}
J(t,r,\theta,\phi)=(-t,r,\pi-\theta,\phi+\pi)
\end{equation}
where $J$ is mapping each exterior region to the other.
While this violates the time-symmetry of the  extended black hole
(for instance, only one smooth constant-time surface exists), on a classical level this cannot be observed locally. 
This means that field theory on the geon is related to field theory on Schwarzschild via the method of images \citep{Smith:2013zqa}.

The Wightman function for the geon can be written  as
\begin{equation}
W_{geon}(x,x') = W_{BH}(x,x')+W_{BH}(x,J(x'))
\end{equation}
in terms of the involution and the black hole Wightman function $W_{BH}(x,x')$.
Therefore, we can explore whether the black hole can be distinguished from its geon counterpart merely by doing some calculations in Schwarzschild. The involution implies that we must consider the Wightman function (i.e. the correlation of the field) \textit{between} the two exterior regions of the extended Kruskal space.
This introduces some difficulties with the notion of `positive frequency'; in fact, it is precisely this tension that introduces time dependence into the Wightman function.

Note that an alternate approach exists:   one might also solve for the modes on the geon directly, picking boundary conditions on the horizon that are invariant under this involution  \citep{Louko:1998dj}. However, since we are interested in the \textit{difference} between the geon and Schwarzschild, it is simpler to take the images approach instead.  This also facilitates our calculation of the transition rate, as we will later show.

We will now briefly review the derivation (see e.g. \citep{Unruh:1976db, Israel:1976ur}) of the Hartle-Hawking vacuum on the Schwarschild spacetime, and point out the differences in the geon case where they occur. In order to solve the Klein-Gordon equation, we can use the ansatz
\begin{equation}
F_{\omega l m}(t,r,\theta,\phi)=\frac{1}{\sqrt{4\pi\omega}}e^{-i\omega t} R_{\omega l}(r)Y_{l m}(\theta,\phi),
\end{equation}
where we suppress the `in' and `up' labels for brevity.
These modes, which are often called Boulware modes, represent positive frequency modes, i.e. particles, with respect to the Schwarzschild coordinate time $t$. However, the vacuum constructed with respect to these modes is ill-behaved in the Schwarzschild spacetime. Specifically, the Boulware vacuum is singular on the horizons of the black hole. The Hartle-Hawking vacuum, on the other hand, is smooth on the horizons, but has non-zero particle content; a stationary observer outside the black hole will observe particles \citep{Unruh:1976db}.

Let us now follow the derivation of the Hartle-Hawking vacuum. Broadly, we construct so-called `Kruskal' modes that annihilate the Hartle-Hawking vacuum, relate them to our Boulware modes, and thus find the Wightman function for the Hartle-Hawking vacuum. Since the Hartle-Hawking vacuum is annihilated by the Kruskal modes, this will allow us to calculate the Wightman function quickly. As above, we will omit `in' and `up' labels, as both types of modes are treated the same way in the Hartle-Hawking vacuum.

First, we note that the Schwarzschild coordinates can also be applied to the parallel exterior  
of the  extended Kruskal space (i.e. the other asymptotically flat region), and therefore the solutions to the Klein-Gordon equation with respect to $(t,r,\theta,\phi)$ may be reused in this region. We thus denote solutions in the parallel exterior as $F^-_{\omega l m}$, and solutions in the physical exterior as $F^+_{\omega l m}$. We will abuse notation, and refer to the extensions to the whole Kruskal space of these modes by the same symbols. $F^+$ thus vanishes in the parallel exterior, and $F^-$ vanishes in the physical exterior. However, since the coordinate time $t$ runs `backwards' in the parallel exterior, i.e. with respect to the global (i.e. Kruskal) time $T$, we associate $F^-$ with the \textit{creation} operator $a^{-\dagger}$ in the parallel exterior, in contrast to $F^+$ and the annihilation operator $a^+$. In terms of these modes, the field operator is
\begin{equation}
\Phi(x)=\sum_{l m} \int_0^{\infty} d\omega (a^+_{\omega l m} F^+_{\omega l m} + a^-_{\omega l m} F^{-*}_{\omega l m} + h.c.)
\end{equation}
Since most calculations involve observers in the physical exterior, the $F^-$ terms usually vanish.

As previously observed  \citep{Israel:1976ur,Unruh:1976db},  we can create modes that have positive frequency with respect to $T$ via a linear superposition of Boulware modes. Using the conventions of \citep{Israel:1976ur}, let us write
\begin{align}
H^+_{\omega l m}&=\frac{1}{\sqrt{2\sinh (4\pi M \omega)}}\left(e^{2\pi M \omega}F^+_{\omega l m}+e^{-2\pi M \omega}F^-_{\omega l m}\right) \\
H^-_{\omega l m}&=\frac{1}{\sqrt{2\sinh (4\pi M \omega)}}\left(e^{2\pi M \omega}F^-_{\omega l m}+e^{-2\pi M \omega}F^+_{\omega l m}\right)
\end{align}
where we use the convention that the positive frequency pair is $(H^+, H^{-*})$.  Note that with respect to $T$,
  $H^+$ has positive frequency whereas $H^-$ has negative frequency. However $H^{-*}$ corresponds to a distinct set of modes from $H^+$. Intuitively, for $\omega$ much larger than the temperature of the black hole, $H^+$ corresponds to modes that are strongly localized in the physical exterior, while $H^{-*}$ corresponds to modes localized in the parallel exterior.
We therefore  associate the annihilation operators $b^+$, $b^-$ with $H^+$, $H^{-*}$ respectively, and the creation operators $b^{+\dagger}$, $b^{-\dagger}$ with $H^{+*}$, $H^-$. In terms of these modes, the field operator may be written
\begin{equation}
\Phi(x)=\sum_{l m} \int_0^{\infty} d\omega (b^+_{\omega l m} H^+_{\omega l m} + b^-_{\omega l m} H^{-*}_{\omega l m} + h.c.)
\end{equation}
This time, both the $H^+$ and $H^-$ parts are non-zero in the physical exterior. However, since the Hartle-Hawking vacuum is annihilated by $b^{\pm}$, we can therefore express the Schwarzschild Wightman function as
\begin{align}
&W_{BH}(x,x')= _{HH}\negthinspace\left\langle 0 |  \Phi(x)\Phi(x') | 0 \right\rangle_{HH}\\
& \quad =\sum_{l m} \int_0^{\infty}d\omega\, (H^+_{\omega l m}(x) H^+_{\omega l m}(x')  + H^{-*}_{\omega l m}(x) H^{-*}_{\omega l m}(x'))
\nonumber
\end{align}

Finally, we express the $H^{\pm}$ functions in terms of the $F^{\pm}$ functions. Consider first the usual case, where $x$ and $x'$ are in the physical exterior, and for convenience $\theta=\theta'=\phi=\phi'=0.$ We then get the usual result
\begin{widetext}
\begin{align}
W_{BH}(x,x')=&\sum_{l m} \int_0^{\infty} \frac{d\omega}{2\sinh(4\pi M\omega)} 
\left(e^{4\pi M\omega} F^+_{\omega l m}(x) F^{+*}_{\omega l m}(x')\right. 
\left.+e^{-4\pi M\omega} F^{+*}_{\omega l m}(x) F^+_{\omega l m}(x')\right) \nonumber\\
=&\sum_{l m} \int_0^{\infty} \frac{d\omega}{8\pi\omega\sinh(4\pi M\omega)} 
\bigg(e^{4\pi M\omega} e^{-i\omega(t-t')}R_{\omega l}(r)R^*_{\omega l}(r')  
Y_{lm}(\theta,\phi)Y^*_{lm}(\theta',\phi')\nonumber\\
&\qquad \qquad  +e^{-4\pi M\omega}e^{+i\omega(t-t')}R^*_{\omega l}(r)R_{\omega l}(r') 
 Y^*_{lm}(\theta,\phi)Y_{lm}(\theta',\phi')\bigg) \nonumber\\
=&\sum_{l m} \int_0^{\infty} \frac{(2l+1)\delta_{m,0}d\omega}{32\pi^2\omega\sinh(4\pi M\omega)} \left(e^{4\pi M\omega} e^{-i\omega(t-t')}R_{\omega l}(r)R^*_{\omega l}(r') 
 + e^{-4\pi M\omega}e^{+i\omega(t-t')}R^*_{\omega l}(r)R_{\omega l}(r')\right)
\end{align}
\end{widetext}
where summation over both `in' and `up' modes is understood as before.
The $\delta_{0,m}$ factor occurs because all higher $m$ spherical harmonics vanish at the pole; therefore, we need only sum over $l$ and $\omega$.

Since we want to calculate the geon Wightman function using the method of images, we also need the case $W(x,J(x'))$ where $J(x')$ is in the parallel exterior. In that case, we get
\begin{widetext}
\begin{align}
W_{BH}(x,J(x'))=&\sum_{l m} \int_0^{\infty} \frac{d\omega}{2\sinh(4\pi M\omega)}
\left(F^+_{\omega l m}(x) F^-_{\omega l m}(J(x')) +F^{+*}_{\omega l m}(x) F^{-*}_{\omega l m}(J(x'))\right)\nonumber\\
=&\sum_{l m} \int_0^{\infty} \frac{d\omega}{8\pi\omega\sinh(4\pi M\omega)} \bigg(e^{-i\omega(t+t')}R_{\omega l}(r)R^*_{\omega l}(r')
 Y_{lm}(\theta,\phi)Y_{lm}(\pi-\theta',\phi'+\pi) \nonumber\\
&\qquad\qquad\qquad +e^{+i\omega(t+t')}R^*_{\omega l}(r)R_{\omega l}(r') Y^*_{lm}(\theta,\phi)Y^*_{lm}(\pi-\theta',\phi'+\pi)\bigg)
\nonumber\\
=&\sum_{l m} \int_0^{\infty} \frac{(-1)^l(2l+1)\delta_{m,0}d\omega}{32\pi^2\omega\sinh(4\pi M\omega)} 
\left(e^{-i\omega(t+t')}R_{\omega l}(r)R^*_{\omega l}(r') + e^{+i\omega(t+t')}R^*_{\omega l}(r)R_{\omega l}(r')\right)
\end{align}
\end{widetext}
Note that the factor $e^{\pm 4 \pi M\omega}$ is no longer present: because we matched $F^+$ with $F^-$, the corresponding factors of $e^{\pm 2 \pi M\omega}$ now cancel, rather than add. This does not remove the exponential suppression characteristic of Hawking radiation: in fact, the resulting expression for the image term decays \textit{faster} than the usual expression for Hawking radiation. However, the magnitude of the image term is now even with respect to $\omega$, and decays to zero even for \textit{negative} $\omega$. This is in contrast to the Hawking radiation expression, which has a term that grows for negative $\omega$.

More specifically, consider a stationary detector, where $r=r'$.
We find that
\begin{align}
W_{BH}(x,x')=&\sum_{l=0}^\infty\int_0^{\infty}\frac{(2l+1)d\omega}{16\pi^2\omega\sinh(4\pi M\omega)}\nonumber\\
&\times\cosh(4\pi M\omega-i\omega(t-t'))|R_{\omega l}(r)|^2,\\
W_{BH}(x,J(x'))=&\sum_{l=0}^\infty \int_0^{\infty} \frac{(-1)^l(2l+1)d\omega}{16\pi^2\omega\sinh(4\pi M\omega)}\nonumber\\
&\times\cos(\omega(t+t'))|R_{\omega l}(r)|^2.
\end{align}
Notice that the image term depends on $t+t'$, not $t-t'$ as one usually sees: that is, the geon vacuum is time-dependent, even in the exterior region.
This implies that $W_{BH}(x,J(x'))$ will be maximized if $t+t'=0$, i.e. the geon contribution is greatest about $t=0$. In some sense, this is as we would expect; the involution $J$ and the correlation of the modes between exterior regions essentially map each point outside the black hole at time $t$ to an antipodal point at time $-t$, so it is natural that the field is most closely correlated when $t'=-t$.

It also appears that higher angular momenta contributions interfere with each other, due to the antipodal nature of $J$. While higher energy modes already are suppressed, this also implies that the geon correlation is suppressed far from the black hole, since in that region many angular momentum values must be considered. This is as we would expect: the nearer the detector is to the black hole, the easier it is to observe its interior structure.

Also interesting is the fact that the \textit{lowest} (absolute) energy modes appear most readily in the geon contribution. This is clearly not a thermal contribution: for instance, it is even in the energy, rather than odd. As we will see later, this implies that the detector must have a small gap and be switched on slowly. In fact, at very low energies, the geon contribution approaches the Schwarzschild contribution in magnitude; this results in a significant enhancement in transition rate.

For completeness, let us also include the Wightman functions for the Unruh vacuum, $W_{BH,U}$ and $W_{J,U}$, when $r=r'$. These differ from the Hartle-Hawking vacuum in that there is no ingoing radiation; therefore, there is no correlation of `in' modes between exteriors, and thus no $W_{J,U}$ contribution. We therefore find that
\begin{align}
W_{BH,U}=&\sum_{l} \int_0^{\infty} \frac{(2l+1)d\omega}{16\pi^2\omega}\nonumber\\
&\times\left(\frac{\cosh(4\pi M\omega - i\omega(t-t'))}{\sinh(4\pi M\omega)}|R^{\textrm{up}}_{\omega l}(r)|^2\right.\nonumber\\
&\left.+e^{-i\omega(t-t')}|R^{\textrm{in}}_{\omega l}(r)|^2\right)\\
W_{J,U}=&\sum_{l} \int_0^{\infty} \frac{(-1)^l(2l+1)d\omega}{16\pi^2\omega}\nonumber\\
&\times\frac{\cos(\omega(t+t'))}{\sinh(4\pi M\omega)}|R^{\textrm{up}}_{\omega l}(r)|^2;
\end{align}

The task that remains is to solve the Klein-Gordon equation on the Schwarzschild spacetime, which we do via refining previous
work along these lines \citep{Hodgkinson:2014iua,Ng:2014kha,Hodgkinson:2013tsa}.
First, if we let $\tilde{R}_{\omega l}=rR_{\omega l}$, the equation for the new radial function becomes a one-dimensional scattering equation,
\begin{equation}
\left(\frac{d^2}{dr^{*2}} + \omega^2 - \left(1-\frac{2M}{r}\right)\left(\frac{l(l+1)}{r^2}+\frac{2M}{r^3}\right)\right)\tilde{R}_{\omega l}=0.
\end{equation}
Therefore, we can use the usual choices of ``in" and ``up" modes, namely
\begin{align}
\tilde{R}^{in}_{\omega l}=&
\begin{cases}
B^{in}_{\omega l}e^{-i\omega r^*}, &r\rightarrow 2M \\
e^{-i\omega r^*} + A^{in}_{\omega l}e^{+i\omega r^*}, &r\rightarrow \infty
\end{cases}\\
\tilde{R}^{up}_{\omega l}=&
\begin{cases}
e^{+i\omega r^*} + A^{up}_{\omega l}e^{-i\omega r^*}, &r\rightarrow 2M\\
B^{up}_{\omega l}e^{+i\omega r^*}, &r\rightarrow \infty
\end{cases}
\end{align}
The reflection and transmission coefficients, $A_{\omega l}$ and $B_{\omega l},$ follow the usual relations for scattering problems.
There is a small difficulty: to normalize these modes, they must be evaluated away from the region in which they are purely exponential. Therefore, we must find appropriate expressions for the modes in their asymptotic regimes, then numerically integrate them to a common point, and compare them. Once this is done, we can normalize them properly.

We will use two expressions from \citep{Leaver:1986} to evaluate our modes in asymptotic regimes, before simply numerically integrating towards a common point; specifically, we can use the Jaff\'{e} expansion near the horizon, and the asymptotic Coulomb expression near infinity, for the unnormalized (i.e. asymptotically $e^{\pm i \omega r^*}$) in and up modes respectively. Letting $2M=1$ for brevity and consistency with \citep{Leaver:1986} we have
\begin{align}
\tilde{R}^{in}_{\omega l}\propto&e^{-i\omega r}(r-1)^{-i\omega}\sum_{n=0}^\infty a_n \left(\frac{r-1}{r}\right)^n \\
\tilde{R}^{up}_{\omega l}\propto&\sqrt{\frac{r}{r-1}}(r-1)^{+i\omega}H^+_{\nu_a}(-\omega,\omega r)e^{-i\theta_{\omega l}}
\end{align}
where $H^+_{\nu_a}(\eta,\omega r)$ is the Coulomb wavefunction of possibly complex order $\nu_a=(-1+\sqrt{(2l+1)^2-12\omega^2})/2$, $\theta_{\omega l}=\omega \log(2\omega)-\pi\nu_a/2-\textrm{ph}\Gamma(\nu_a+1-i\omega)$ is a phase shift (which Leaver omits) to match $e^{+i\omega r^*}$, and the Jaff\'{e} coefficients $a_n$ are generated by the recurrence relation
\begin{align}
a_{-1}=&0,\nonumber\\
a_0=&1,\nonumber\\
0=&(1+n)(1+n-2i\omega)a_{n+1}\nonumber\\
&+(-1-l(l+1)-2n(1+n))a_n+n^2 a_{n-1}.
\end{align}
While this requires some numerical precision, the fact that geon contributions are concentrated at low energy means that we do not have to worry about loss of precision due to oscillation.

\section{Switching the detector}

The time dependence of the geon means that the switching function we use is important. It is not obvious, for instance, what the effect of time translation on the response function is. In the spirit of \citep{Ng:2016hzn} we will express the transition function in terms of the Fourier-transformed switching function, by treating the $\tau,\tau'$ integrals as Fourier transforms. \tcb{Here we will use the usual convention for quantum physics, namely}
\begin{align}
\hat{\chi}(\tilde{\omega})=&\frac{1}{\sqrt{2\pi}}\int_{-\infty}^{\infty}e^{-i\tilde{\omega}\tau}\chi(\tau).
\end{align}

The presence of the ``negative frequency" part of the Wightman function means that our expression becomes somewhat more complicated, even in the Schwarzschild case. Modulo the usual caveats about smooth switching and regularization, we can show that
\begin{align}
F_{BH}(\Omega)=&\sum_{l=0}^\infty \int_{0}^\infty \frac{(2l+1)d\tilde{\omega}}{16\pi\omega\sinh(2\pi\omega)}
\left(|R^{\textrm{in}}_{\omega l}|^2+|R^{\textrm{up}}_{\omega l}|^2\right)\nonumber\\
&\times\left(e^{-2\pi\omega}|\hat{\chi}(\tilde{\omega}-\Omega)|^2+e^{+2\pi\omega}|\hat{\chi}(\tilde{\omega}+\Omega)|^2\right),\\
F_{J}(\Omega)=&\sum_{l=0}^\infty \int_{0}^\infty \frac{(-1)^l(2l+1)d\tilde{\omega}}{16\pi\omega\sinh(2\pi\omega)}
\left(|R^{\textrm{in}}_{\omega l}|^2+|R^{\textrm{up}}_{\omega l}|^2\right)\nonumber\\
&\times 2\Re\left(\hat{\chi}(\tilde{\omega}-\Omega)\hat{\chi}(\tilde{\omega}+\Omega)\right),
\end{align}
where $\tilde{\omega}=\omega/\sqrt{(1-1/r)}$ is the `local' frequency of the mode as measured by the detector. In this paper, we will use $F_J$, or ``geon contribution", to refer to the $W(x,J(x'))$ part of the response function; this is to avoid confusion with the total response of a detector above a geon, which would be $F_{geon}=F_{BH}+F_J.$

As mentioned earlier, the geon contribution is time-dependent. Therefore, when expressed in terms of Fourier transforms, the transition function depends in part on the complex phases of the modes and the (Fourier space) switching function; from our discussion of the Wightman function, we can conclude that the contribution will be greatest when the switching function is centered about zero. More specifically, let $\xi(\tau)=\chi(\tau-\tau_0).$ Then, since $\hat\xi(\tilde\omega)=e^{-i\tilde\omega \tau_0}\hat\chi(\tilde\omega),$ one finds that the time-translated switching function has a corresponding geonic response
\begin{align}
F_{J,\tau_0}(\Omega)=&\sum_{l=0}^\infty \int_{0}^\infty \frac{(-1)^l(2l+1)d\tilde{\omega}}{16\pi\omega\sinh(2\pi\omega)}
\left(|R^{\textrm{in}}_{\omega l}|^2+|R^{\textrm{up}}_{\omega l}|^2\right)\nonumber\\
&\times 2\Re\left(e^{-2i\tilde\omega\tau_0}\hat{\chi}(\tilde{\omega}-\Omega)\hat{\chi}(\tilde{\omega}+\Omega)\right),
\end{align}
which demonstrates the de-phasing effect to which we previously alluded.

As well, the geon contribution is largest near zero detector gap. However, analysis of the low-energy limit of the equations is challenging; while \citep{Leaver:1986} attests that the Coulomb asymptotic expression is exact in the limit $\omega \rightarrow 0$, the peculiar combination of limits we must take to find the correct \textit{normalization} of the modes makes finding an analytical expression challenging. This may \textit{a priori} raise certain issues of numerical accuracy, since the integrand of $F$ will diverge unless the modes approach zero in a particular way. However, since we know from \citep{Louko:2007mu} that these expressions must be infrared-safe, we will take the na\"ive approach and introduce a numerical infrared cutoff. (Of course, ultraviolet safety is ensured by our choice of a smooth switching function \citep{Louko:2007mu}.)

For  simplicity we will work with Gaussian functions. As in the shell case, a question arises of whether the use of switching functions of noncompact support is justified; specifically, whether it makes sense to speak of a ``measurement" that continues eternally. However, since the geon effect is concentrated at low frequency, the use of compactly supported switching functions (i.e. measurements in finite time) is unlikely to make much difference. (In fact, we have done preliminary calculations demonstrating that compact support makes little difference.)

In order to illuminate the previous discussions, let us consider the particular switching function \mbox{$\chi(\tau)=e^{-\tau^2/2\sigma^2}$}, where $\sigma$ is the width of the proper-time switching function. The Fourier transform is then $\hat\chi(\tilde\omega)=\sigma e^{-\sigma^2\tilde\omega^2/2}$. In that case, the response functions become  
\begin{widetext}
\begin{align}
F_{BH}(\Omega)=&\sum_{l=0}^\infty \int_{0}^\infty \frac{(2l+1)d\tilde{\omega}}{16\pi\omega\sinh(2\pi\omega)}
\left(|R^{\textrm{in}}_{\omega l}|^2+|R^{\textrm{up}}_{\omega l}|^2\right) \sigma \left(e^{-2\pi\omega-\sigma^2(\tilde\omega-\Omega)^2}+e^{+2\pi\omega-\sigma^2(\tilde\omega+\Omega)^2}\right)\nonumber\\
=&\sum_{l=0}^\infty \int_{0}^\infty \frac{(2l+1)d\tilde{\omega}}{8\pi\omega\sinh(2\pi\omega)}
\left(|R^{\textrm{in}}_{\omega l}|^2+|R^{\textrm{up}}_{\omega l}|^2\right) \sigma e^{-\sigma^2(\tilde\omega^2+\Omega^2)}\cosh(2\pi\omega-2\sigma^2\tilde\omega\Omega)\\
F_{J}(\Omega)=&2\sum_{l=0}^\infty \int_{0}^\infty \frac{(-1)^l(2l+1)d\tilde{\omega}}{16\pi\omega\sinh(2\pi\omega)}
\left(|R^{\textrm{in}}_{\omega l}|^2+|R^{\textrm{up}}_{\omega l}|^2\right) \sigma e^{-\frac{\sigma^2}{2}\left((\tilde\omega-\Omega)^2+(\tilde\omega+\Omega)^2\right)}\nonumber\\
=&\sum_{l=0}^\infty \int_{0}^\infty \frac{(-1)^l(2l+1)d\tilde{\omega}}{8\pi\omega\sinh(2\pi\omega)}
\left(|R^{\textrm{in}}_{\omega l}|^2+|R^{\textrm{up}}_{\omega l}|^2\right) \sigma e^{-\sigma^2(\tilde\omega^2+\Omega^2)}\nonumber\\
=&e^{-\sigma^2\Omega^2}F_J(0)
\end{align}
and the effect of time translation on $F_J$ is
\begin{align}
F_{J,\tau_0}(\Omega)=&\sum_{l=0}^\infty \int_{0}^\infty \frac{(-1)^l(2l+1)d\tilde{\omega}}{16\pi\omega\sinh(2\pi\omega)}
\left(|R^{\textrm{in}}_{\omega l}|^2+|R^{\textrm{up}}_{\omega l}|^2\right) \times 2\sigma e^{-\sigma^2(\tilde\omega^2+\Omega^2)}\cos(2\tilde\omega\tau_0)\, .
\end{align}
\end{widetext}
Remarkably, it appears that the geon contribution $F_J(\Omega)$ depends on the gap $\Omega$ in an almost trivial way. It is also notable that if $\sigma$ is selected such that only very small $\tilde\omega$ (and thus only $l=0$) contribute, then we find that $F_{J}(0)/F_BH(0)\approx 1.$ In other words, we find that the geon signal can equal the ``background'' in magnitude, in the slow-switching zero-gap limit. Thus, we have confirmed that the geon and the Schwarzschild black hole can, in principle, be distinguished through the response of particle detectors.

One may prefer the Unruh vacuum on physical grounds. In that case, one simply drops the `in' terms from the summation for $F_J$. The effect on $F_{BH,U}$ is more complicated:
\begin{widetext}
\begin{align}
F_{BH,U}=&\sum_{l=0}^\infty \int_{0}^\infty \frac{(2l+1)d\tilde{\omega}}{8\pi\omega}\sigma\left(e^{-\sigma^2(\tilde\omega-\Omega)^2}|R^{\textrm{in}}_{\omega l}|^2 + e^{-\sigma(\tilde\omega^2+\Omega^2)}\frac{\cosh(2\pi\omega-2\sigma\tilde\omega\Omega)}{\sinh(2\pi\omega)}|R^{\textrm{up}}_{\omega l}|^2\right) \, .
\end{align}
\end{widetext}
We note here that there are questions of how scaling different parts of the switching function (i.e. choosing non-Gaussianities) affects the final results; while we will not discuss them further, an interested reader may refer to \citep{Fewster:2016ewy}, which reached this formalism independently, and investigates these issues with great rigor.

\section{Transition rates}

In order to compare our results with those of the BTZ case \citep{Smith:2013zqa}, we may also consider \textit{sudden} switching, in order to yield a transition rate.  Although sudden switching causes a number of divergences in the transition function, these divergences are independent of the field \citep{Louko:2007mu}; therefore, by taking a difference between fields, we can find a well-behaved expression for the transition rate. Our expression of the geon transition function in terms of $W_J$ is well-suited for this;  the ultraviolet divergences that usually result from such a calculation cancel out. As well, since the black hole vacuum is time-independent, we can also use the na\"ive expression for the transition rate of the detector in a Schwarzschild spacetime. (The latter was previously calculated by \citep{Hodgkinson:2013tsa}, and serves as a check on our methods.)

Employing the sudden switching function $\chi_{\tau_0}(\tau)=\Theta(\tau_0-\tau)$, the Fourier transform becomes 
$$\hat\chi_{\tau_0}(\tilde\omega)=\sqrt{\frac{\pi}{2}}\delta(\tilde\omega)-\frac{i e^{i \tau_0 \tilde\omega}}{\sqrt{2\pi}\tilde\omega}$$
and so the dependence of $W_J$ on the switching function is
\begin{align}
&\tilde\chi_{\tau_0}(\tilde\omega-\Omega)\tilde\chi_{\tau_0}(\tilde\omega+\Omega) = -\frac{e^{2i\tau_0 \tilde\omega}}{2\pi(\tilde\omega^2-\Omega^2)} \\
&\qquad \qquad -\frac{i e^{i\tau_0(\tilde\omega-\Omega)}}{2(\tilde\omega-\Omega)}\delta(\tilde\omega+\Omega) -\frac{i e^{i\tau_0(\tilde\omega+\Omega)}}{2(\tilde\omega+\Omega)}\delta(\tilde\omega-\Omega)  \nonumber
\end{align}
for $\Omega \neq 0$.
Since we have restricted $\tilde\omega>0$, and we only require the real part, we can further simplify this to
\begin{align}
&\partial_{\tau_0}2\Re\left(\tilde\chi_{\tau_0}(\tilde\omega-\Omega)\tilde\chi_{\tau_0}(\tilde\omega+\Omega)\right) \nonumber\\
&= \cos(2\tau_0 \Omega)\delta(\tilde\omega-|\Omega|) +\frac{2\tilde\omega\sin(2\tau_0\tilde\omega)}{\pi(\tilde\omega^2-\Omega^2)}.
\end{align}
It is then simple to take the time derivative and thus find the transition rate. Note that the `singular' last term should be understood in the Cauchy principal value sense; this is well-defined as a distribution, and is additionally the limit of smooth switching as switching time approaches zero. As well, the assumption that $\Omega \neq 0$ is only used to simplify the algebra; in principle, we could also set $\Omega = 0$, then use the time derivative to remove the $\delta(\tilde\omega)^2$ term.

We next consider $|\tilde\chi(\tilde\omega-\Omega)|^2$
\begin{align}
|\tilde\chi_{\tau_0}(\tilde\omega-\Omega)|^2=&\frac{\pi}{2}\delta(\tilde\omega-\Omega)^2-\frac{i e^{i\tau_0(\tilde\omega-\Omega)}}{2(\tilde\omega-\Omega)}\delta(\tilde\omega-\Omega)\nonumber\\
&+\frac{i e^{-i\tau_0(\tilde\omega-\Omega)}}{2(\tilde\omega-\Omega)}\delta(\tilde\omega-\Omega)
+\frac{1}{2\pi\tilde\omega}
\end{align}
 which is required for $W_{BH}$.
Since we wish to find a transition \textit{rate}, we take the time derivative to find
\begin{align}
\partial_{\tau_0}|\tilde\chi_{\tau_0}(\tilde\omega-\Omega)|^2=&e^{i\tau_0 (\tilde\omega-\Omega)}\delta(\tilde\omega-\Omega)= \delta(\tilde\omega-\Omega),
\end{align}
and similarly $\partial_{\tau_0}|\tilde\chi_{\tau_0}(\tilde\omega+\Omega)|^2=\delta(\tilde\omega+\Omega)$. Hence the dependence of $\dot F_{BH}(\Omega)$ on the switching function (where $\tilde\omega>0$) is
\begin{align}
\partial_{\tau_0}&\left(e^{-2\pi\omega}|\tilde\chi_{\tau_0}(\tilde\omega-\Omega)|^2+e^{+2\pi\omega}|\tilde\chi_{\tau_0}(\tilde\omega+\Omega)|\right)\nonumber\\
=&e^{-\Omega/2T_{loc}}\delta(\tilde\omega-|\Omega|)
\end{align}
where $T_{loc}=T_H/\sqrt{f(r)}=1/(4\pi\sqrt{1-1/r})$.

Substituting these expressions for the Fourier transformed switching function, we find that the transition rates are
\begin{align}
\dot F_{BH}(\Omega)=&
\sum_{l=0}^\infty  \frac{(2l+1)e^{-2\pi\tilde\Omega}}{16\pi\tilde\Omega\sinh(2\pi\tilde\Omega)}(|R^{in}_{\tilde\Omega l}|^2+|R^{up}_{\tilde\Omega l}|^2)\nonumber\\
\dot F_J(\Omega)=
&\sum_{l=0}^\infty  \frac{(-1)^l (2l+1)\cos(2\tau_0 \Omega)}{16\pi\tilde\Omega\sinh(2\pi\tilde\Omega)}(|R^{in}_{\tilde\Omega l}|^2+|R^{up}_{\tilde\Omega l}|^2)\nonumber\\
&+\sum_{l=0}^\infty \int_{0}^\infty \frac{(-1)^l (2l+1)d\tilde\omega}{16\pi \omega \sinh(2\pi\omega)}\left(|R^{\textrm{in}}_{\omega l}|^2+|R^{\textrm{up}}_{\omega l}|^2\right)\nonumber\\
&\qquad \times \frac{2\tilde\omega\sin(2\tau_0\tilde\omega)}{\pi(\tilde\omega^2-\Omega^2)}
\end{align}
where $\tilde\Omega=\Omega\sqrt{f(r)}$ is not the proper detector gap, but the detector gap as viewed by an observer at infinity.

\section{Results}

The introduction of a time-dependence in the vacuum of the geon introduces a wealth of parameters to examine, compared to the Schwarzschild case. Some relevant parameters include the gap of the detector, the choice of vacuum, the distance of the detector from the black hole, the width of the switching function, and the \textit{time translation} of the switching function.

Since we have shown that the geon contribution to the response is concentrated at low energy, we will adjust most of our parameters accordingly, e.g. keeping the gap small (even zero), placing the detector near the horizon, switching over long periods of time, and switching around $t=0,$ varying individual parameters as necessary. Specifically, unless otherwise specified, let us pick a gap of $\Omega=0$, $r=3r_S$, $\sigma=100r_S/c$, $\tau_0=0$.  Note that such a long switching time, while ``unnatural'', is operationally plausible for small black holes: for instance, a black hole of ten solar masses would correspond to a switching time of 10 milliseconds.

\begin{figure}[hbtp]
\centering
\includegraphics[width=0.45\textwidth]{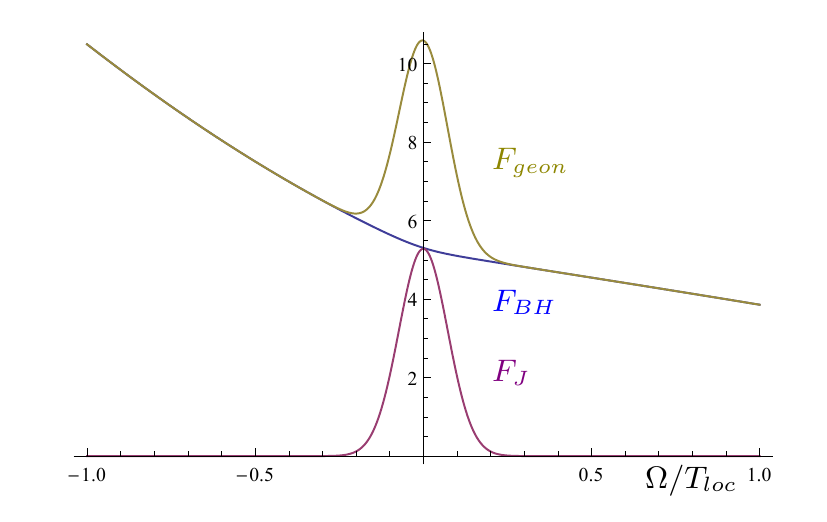}
\caption{Response functions for black hole (blue), geon part (red), and geon total (green) versus $\Omega$. 
}
\label{gapgraph}
\end{figure}

First, let us consider the gap of the detector. It is immediately apparent that the Schwarzschild response is the response found in \citep{Hodgkinson:2013tsa}, convolved with a Gaussian; our choice of $\sigma$ results in a very narrow Gaussian, and so our black hole response is very similar to those found in \citep{Hodgkinson:2013tsa}, up to normalization. (Specifically, the normalization of the Fourier transform of $\delta(k)$.)
Figure \ref{gapgraph} shows that for very small gap, the geon contribution approaches the Schwarzschild ``background''; in other words, the total response of such a detector above a geon is almost twice that of a detector above a traditional black hole. However, the geon contribution decays rapidly away from zero: specifically, the geon contribution $F_J$ decays in a Gaussian manner, while the Schwarzschild part $F_{BH}$ decays in a merely exponential fashion. This induces a characteristic exponential-plus-Gaussian shape in the total geon response $F_{geon}=F_J+F_{BH}$.

Noting earlier explorations of the effect on the Schwarzschild response \citep{Hodgkinson:2013tsa}, we now consider the effect of different vacua on the geon contribution.  While the Unruh part of the geon contribution was smaller, the relative magnitude of the geon contribution against the Schwarzschild contribution, $W_J/W_{BH}$, is similar at $r=3r_S$. 

\begin{figure}[hbtp]
\centering
\includegraphics[width=0.45\textwidth]{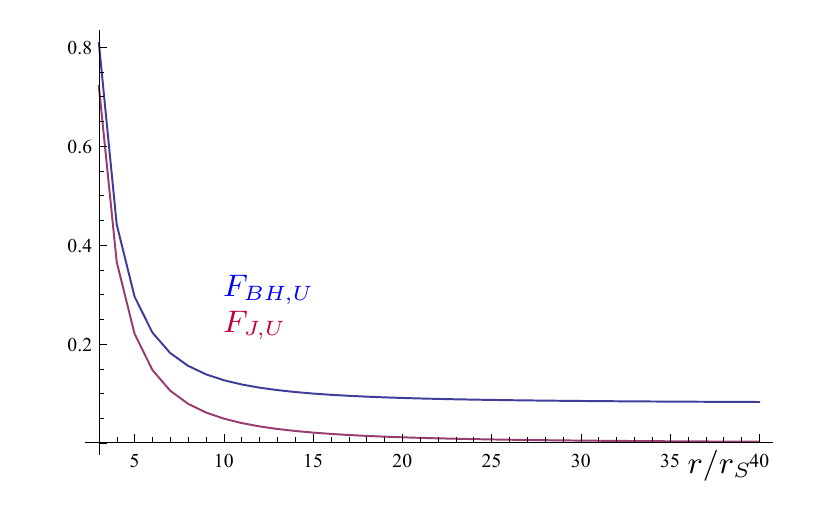}
\caption{Response functions for varying detector radius (Unruh vacuum).}
\label{radiusugraph}
\end{figure}

Next, we consider the effect of the distance of the detector from the geon, plotted below in figure \ref{radiushhgraph}. It is immediately clear that the Hartle-Hawking vacuum has a rather undesirable consequence: namely, the infalling radiation causes a (Schwarzschild) detector response far away from the black hole, which is not what we would expect of a physical black hole. Using the Unruh vacuum removes this irregularity, as seen in figure \ref{radiusugraph}. However, the geon part of the response does decay with distance, as expected (albeit rather gradually in the Hartle-Hawking case).

\begin{figure}[hbtp]
\centering
\includegraphics[width=0.45\textwidth]{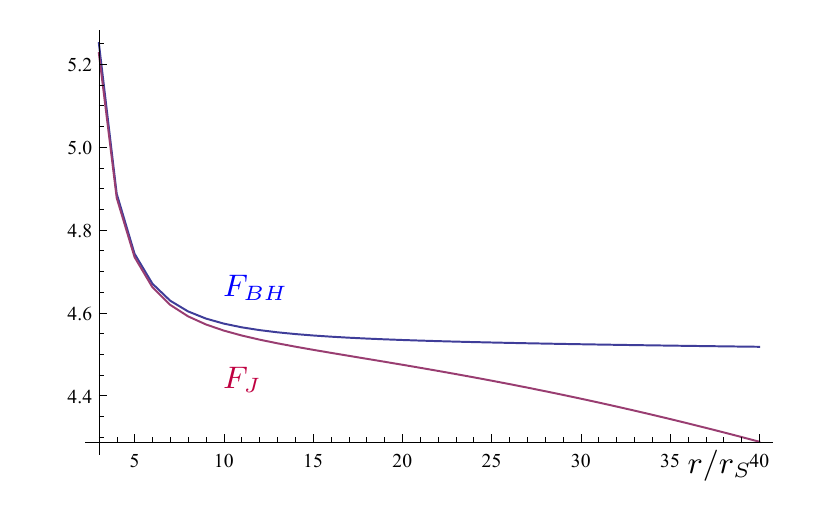}
\caption{Response functions for varying detector radius (Hartle-Hawking vacuum).
}
\label{radiushhgraph}
\end{figure}

\begin{figure}[hbtp]
\centering
\includegraphics[width=0.45\textwidth]{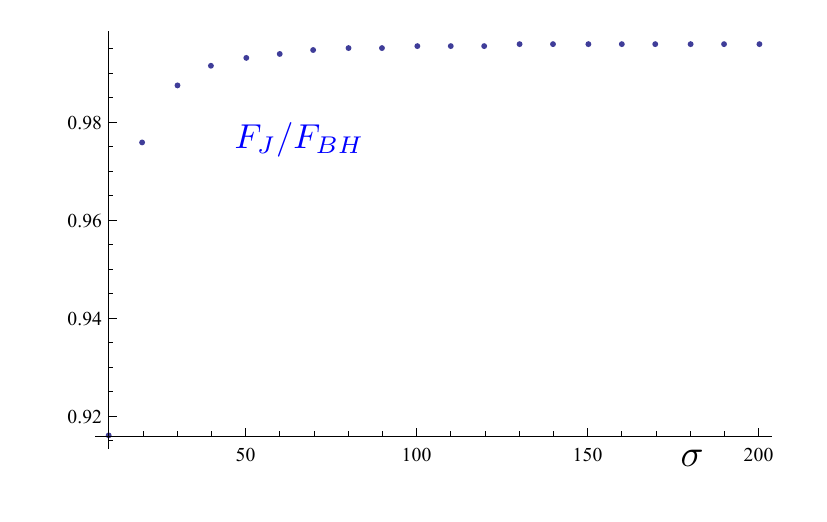}
\caption{Geon relative signal for varying switching width $\sigma$.}
\label{widthgraph}
\end{figure}

As mentioned previously, the low-energy nature of the geon response suggests that we must use slow switching to detect it; as previous experience with time-dependent switching functions has shown \citep{Ng:2016hzn},   faster switching causes a broader dependence on field mode frequency, which may mask the effect we wish to observe. Numerical calculations bear this out; although there is only slight decay in the signal, $F_J/F_{BH}$, it is present in figure \ref{widthgraph}.

\begin{figure}[hbtp]
\centering
\includegraphics[width=0.5\textwidth]{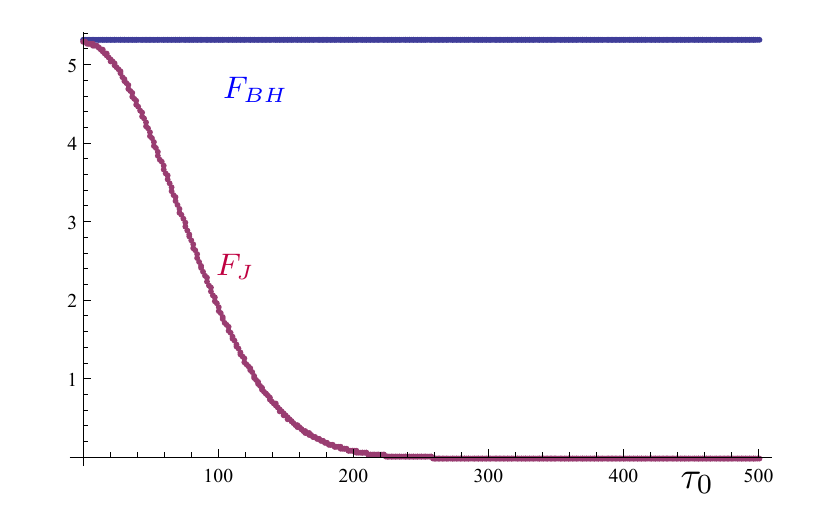}
\caption{Response functions for black hole (blue) and geon part (red) versus time translation.}
\label{translationgraph}
\end{figure}

We may also consider the effect of time-translation on the response. Since the Schwarzschild black hole is time-translation invariant, the response does not depend on time; however, the geon contribution experiences `de-phasing' as a result. As we would expect, the total geon contribution is consequently decreased. However, since this de-phasing causes oscillation in the integral, large time-translations are beyond our numerical simulation. The implications of figure \ref{translationgraph} on geon identification `in the wild' are rather sobering: the effect we have observed will not be observable unless we are present during the moment $t=0$ of greatest correlation.

\begin{figure}[hbtp]
\centering
\includegraphics[width=0.5\textwidth]{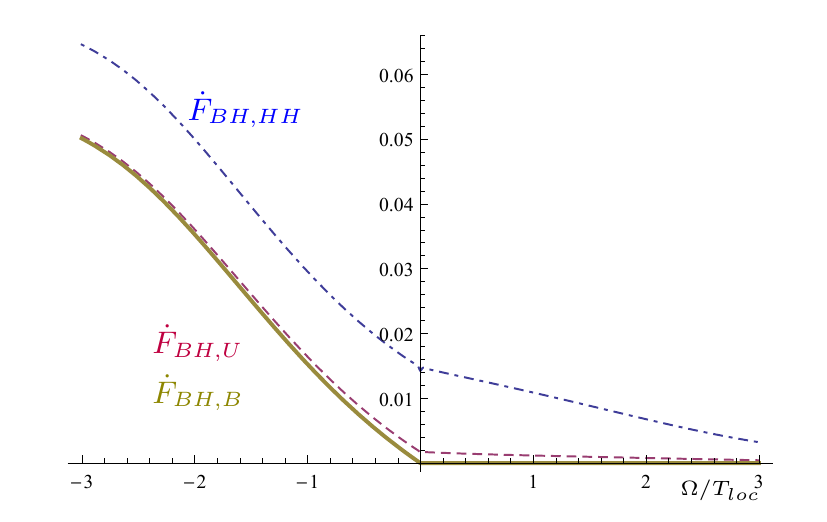}
\caption{Transition rates for black hole versus gap, for the Hartle-Hawking (dot-dashed), Unruh (dashed), and Boulware (solid) vacuum states.}
\label{bhrategraph}
\end{figure}

As a test for the transition rate formalism, we computed the transition rate for the Schwarzschild black hole as a function of $\Omega$, seen in figure \ref{bhrategraph}. The three curves correspond to the Hartle-Hawking, Unruh, and Boulware vacua. Of course, as all three vacuum states are time-independent, the transition rates are constant in time. These results are identical to those of \citep{Hodgkinson:2013tsa}.

\begin{figure}[hbtp]
\centering
\includegraphics[width=0.5\textwidth]{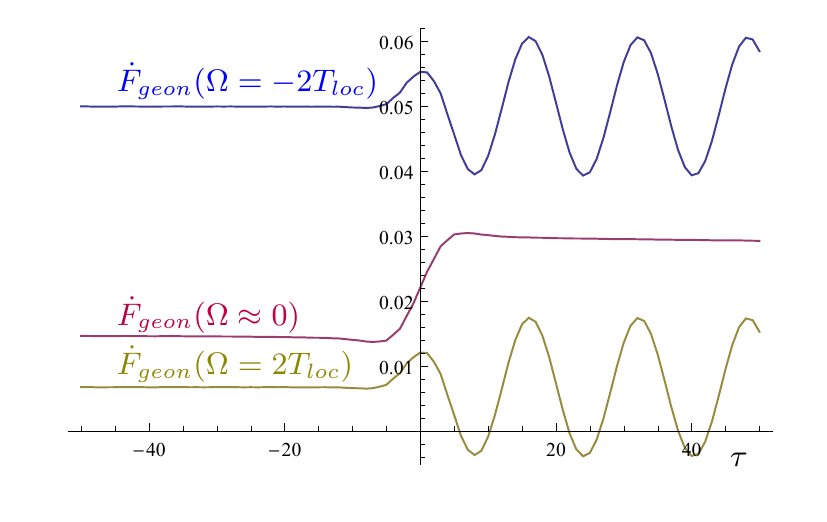}
\caption{Transition rate for $\Omega/T_{loc}$ = -2 (top), 0 (middle), 2 (bottom)}
\label{rategraph}
\end{figure}

Finally we compute the transition rates of the geon for $\Omega/T_{loc}=-2,0.01,2.$ and display the results 
in figure \ref{rategraph}.  The resemblance with the BTZ case  \citep{Smith:2013zqa} is striking.
 As one might expect, since there is little correlation before $\tau=0,$ the geon contribution is weak there; however, for $\tau>0$, the $\tau'=-\tau$ correlation switches on, and thus a time-dependent signal arises.  As Hodgkinson notes \citep{Hodgkinson:2013tsa}, however, this definition of `transition rate' requires that the switch-on of the detector occurs in the distant past; any experimental proposal might be better modelled by the finite-time formalism. 

\section{Conclusion}

Classically, the Schwarzschild black hole and the $\mathbb{RP}^3$ geon are locally identical, where `local' includes the entire exterior of the black hole. It might reasonable to expect that a quantum field in either scenario should be indistinguishable, especially if it is in a vacuum state. On the other hand, the vacuum state of a field in a black hole background contains local excitations, so we might instead expect that a detector may observe some difference in its counting statistics. Our calculations support the latter view: an Unruh-DeWitt detector can distinguish the vacua of quantum fields of a geon from that of a regular black hole.


Our findings are in agreement with previous work on UDW detectors in the vicinity of black holes and geons. The expression we found for the black hole transition rate matches that which is found in \citep{Hodgkinson:2013tsa}. As well, we find that the qualitative features suggested in \citep{Louko:1998dj}, and computed in 2+1 dimensions in \citep{Smith:2013zqa}, are also exhibited for the Schwarzschild black hole and $\mathbb{RP}^3$ geon.  Furthermore, our calculations on Gaussian switching represent a more operationally accessible regime, allowing some quantification of how sensitive a physical experiment would need to be.

The limitations of this finding must also be understood. Firstly, as our calculations show, the geon response is on the same energetic scale as the black hole response, and perhaps even smaller. Any `real-world' measurement will require a correspondingly sensitive detector. Second, as was suggested in \citep{Louko:1998dj} and \citep{Smith:2013zqa}, the geon response is greatly suppressed away from the `critical moment', suggesting that an unlucky observer might still observe no difference.  Indeed, if the detector is not present and active at the time $\tau=0$, the geon signal is so greatly suppressed as to be virtually unobservable.

On a more fundamental level, we have not escaped the laws of causality, nor of the strict version of
topological censorship: the laws of quantum field theory still prevent any \textit{spacelike} change in spacetime from being observable, 
and we have not claimed to have found any active probing (e.g. traversal) of a topological feature, as forbidden by 
the topological censorship theorem \citep{Friedman:1993ty}.  However, we have demonstrated in concrete terms that \textit{passive} topological censorship does not hold, and that a UDW detector outside of an event horizon has a response that is sensitive to the interior topology.
 
In conjunction with earlier work \cite{Smith:2013zqa,Martin-Martinez:2015qwa,Ng:2016hzn}, our results indicate that  the vacuum carries some information about the global features of spacetime. It would be interesting to better understand the implications of this for other quantum information tasks such as entanglement harvesting.

\section*{Acknowledgements}
This work was supported in part by the Natural Science and Engineering Research Council of Canada. E. M-M. acknowledges funding of the Ontario Early Researcher Award.

\bibliography{inthegeon}

\end{document}